\documentclass[%
 reprint,
 superscriptaddress,
 amsmath,amssymb,
 aps,
]{revtex4-2}

\usepackage{graphicx}% Include figure files
\usepackage{dcolumn}% Align table columns on decimal point
\usepackage{bm}% bold math
\begin{document}

\preprint{APS/123-QED}

\title{Measurement of Doppler effects in cryogenic buffer-gas cell}% Force line breaks with \\

\author{Ayami Hiramoto}
 \affiliation{Research Institute for Interdisciplinary Science, Okayama University, Okayama 700-8530, Japan}%
\author{Masaaki Baba}
\affiliation{Department of Chemistry, Graduate School of Science, Kyoto University, Kyoto 606-8502, Japan}%
\affiliation{Molecular Photoscience Research Center, Kobe University, Kobe 657-8501, Japan}
\author{Katsunari Enomoto}
\affiliation{Department of Physics, University of Toyama, Toyama 930-8555, Japan}
\author{Kana Iwakuni}
\affiliation{Institute for Laser Science, University of Electro-Communications, 1-5-1 Chofugaoka, Chofu, Tokyo 182-8585, Japan}
\author{Susumu Kuma}
\affiliation{Atomic, Molecular and Optical Physics Laboratory, RIKEN, 2-1 Hirosawa, Wako, Saitama 351-0198, Japan}
\author{Yuiki Takahashi}
\affiliation{Division of Physics, Mathematics, and Astronomy, California Institute of Technology, Pasadena, CA 91125, USA}
\author{Reo Tobaru}
\affiliation{Research Institute for Interdisciplinary Science, Okayama University, Okayama 700-8530, Japan}
\author{Yuki Miyamoto}%
 \email{miyamo-y@cc.okayama-u.ac.jp }
\affiliation{Research Institute for Interdisciplinary Science, Okayama University, Okayama 700-8530, Japan}%

\date{\today}% It is always \today, today,

\begin{abstract}
Buffer-gas cooling is a universal cooling technique for molecules and used for various purposes. One of its applications is using molecules inside a buffer-gas cell for low-temperature spectroscopy. Although a high-intensity signal is expected in the cell, complex molecular dynamics is a drawback for precise spectroscopy. In this study, we performed high-resolution absorption spectroscopy of low-J transitions in the $\tilde{A}^2\Pi (0,0,0)-\tilde{X}^2\Sigma ^+(0,0,0)$ band of calcium monohydroxide (CaOH). CaOH molecules were produced by laser ablation in a copper cell and cooled to $\sim$5\,K using helium buffer gas. We probed the Doppler effects in a buffer-gas cell by injecting counter-propagating lasers inside the cell. The time evolutions of the Doppler width and shift were simulated using a dedicated Monte Carlo simulation and compared with data.

\end{abstract}

\maketitle

%\tableofcontents

\section{Introduction}

Cold molecules are used in wide-ranging sciences, such as cold chemistry~\cite{Ospelkaus2010,Balakrishnan2016}, high-precision measurements in fundamental physics~\cite{Safronova2018,DeMille2017,Cairncross2019}, and also proposed to quantum computing~\cite{DeMille2002,Sawant2020}. While several possible methods generate cold molecules, buffer-gas cooling is a universal technique for producing high-density cold molecules, that can be applied to many species regardless of their properties~\cite{Patterson2010,Hutzler2012}. In this method, hot target molecules are introduced in a cryogenically cooled cell and thermalized by collision with ultracold buffer gas, such as helium and neon. The molecules are cooled down to a few Kelvin and are usually extracted from the aperture of the cell. The extracted molecules are used directly as a beam~\cite{Santamaria2016,acme2018} or can be further cooled for trapping by laser cooling~\cite{Shuman2010,Vilas2022,Zhang2022} and other techniques~\cite{Stuhl2012,Augenbraun2021}.

One application of buffer-gas cooling is to use the produced molecules for low-temperature spectroscopy~\cite{Santamaria2015,Spaun2016,Iwata2017}. Cooling down molecules to ultracold temperatures reduces the Doppler width and enable high-resolution spectroscopy. Molecules inside and outside a cell can be used. However, the major advantage of probing molecules inside a cell in spectroscopy is the generation of high-intensity signals owing to higher molecule density inside a cell than that outside it. Thus far, spectroscopies of calcium monohydroxide molecules ($^{40}$Ca$^{16}$O$^{1}$H)~\cite{Takahashi2022} and free-base phthalocyanin~\cite{Miyamoto2022} have been conducted inside a cell. In these studies, helium was used as the buffer gas, and the target molecules were provided by laser ablation. Although a higher molecular density is expected in a cell, one drawback is the complex dynamics possibly caused by the helium flow after it thermalizes the target molecules. The complex dynamics affects observed spectra, resulting in a systematic uncertainty as observed in Ref.~\cite{Takahashi2022}. Therefore, careful treatment of this effect is required for precision spectroscopy. Ref.~\cite{Skoff2011} reported the time evolutions of the rotational temperature and Doppler width of YbF in a buffer gas cell. In this reference, the mechanism of the time evolution was discussed and the importance of heating of helium was highlighted.

In this study, we performed high-resolution absorption spectroscopy of low-J transitions in the $\tilde{A}^2\Pi (0,0,0)-\tilde{X}^2\Sigma ^+(0,0,0)$ band of calcium monohydroxid (CaOH) in a buffer-gas cell. Absorption spectra were obtained by injecting counter-propagating lasers inside the cell. Millisecond-scale time evolutions of both the Doppler width and shift were observed in these spectra, because of the importance of the Doppler shift in spectroscopy. In addition, we performed a Monte Carlo simulation of CaOH cooling based on the thermalization model proposed in Ref.~\cite{Skoff2011}. The Doppler width was reproduced well, whereas the shift could not. These results suggested complex helium flows in buffer gas cells.

\section{Experimental setup}

\begin{figure}[h]
\centering
\includegraphics[height=8.0cm]{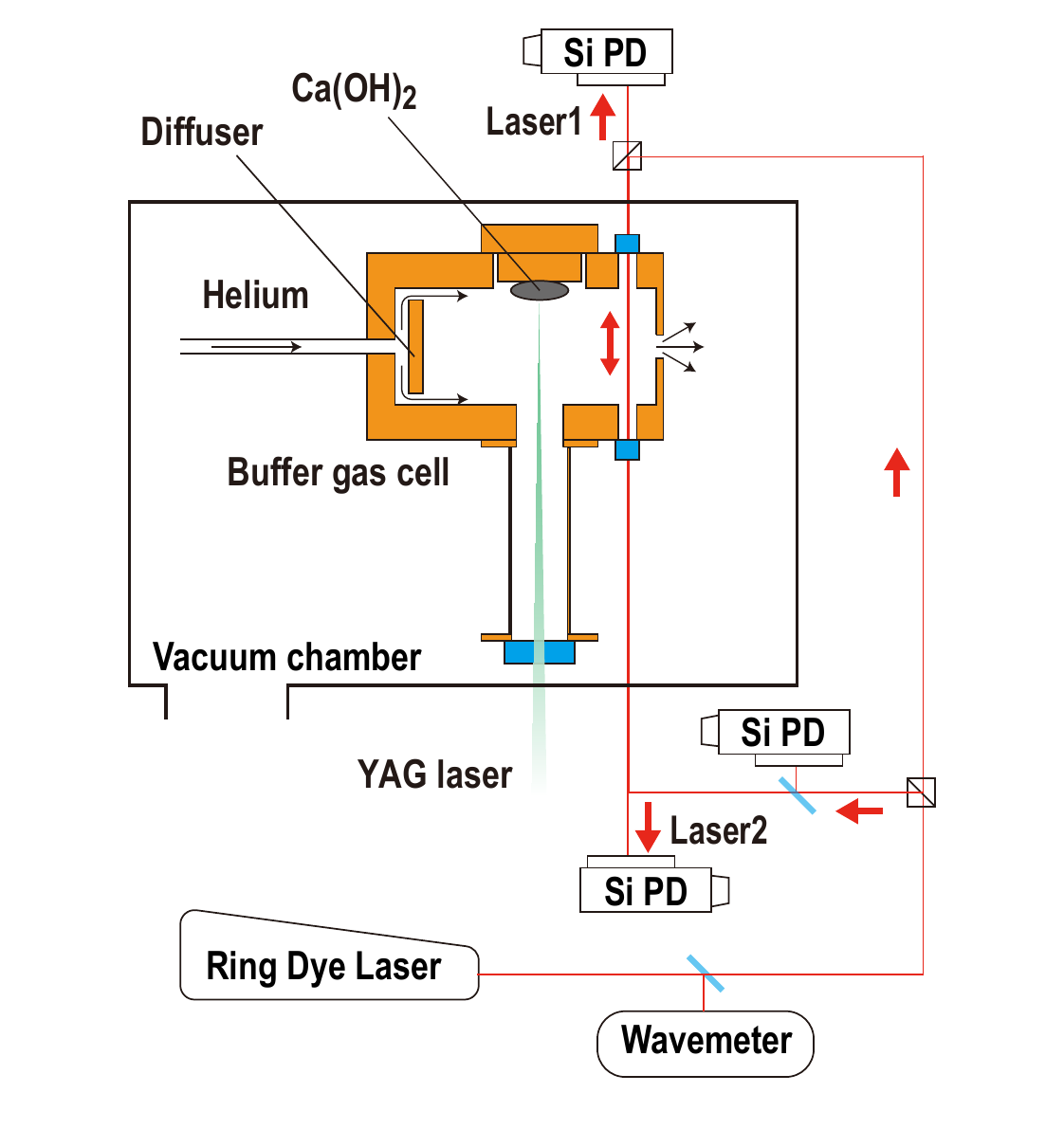}
\caption{\label{fig:setup}Experimental setup: Ablated CaOH molecules are cooled by collision with helium buffer gas in cylindrical copper cell and extracted through aperture. Counter-propagating lasers are injected inside cell perpendicular to expected molecular flow.}
\end{figure}

The buffer-gas cell used in this study was the same as used in Ref.~\cite{Takahashi2022}, to which we added counter-propagating probe lasers. Figure~\ref{fig:setup} shows our measurement setup. The buffer-gas cell is made of a copper block and has a cylindrical cavity that is 5-cm long and 2.5\,cm in diameter. The cell is attached to a 4\,K stage of a pulse tube refrigerator (Sumitomo Heavy Industies SRP-062B) and held at $\sim$5\,K. Helium buffer gas is introduced from an inlet tube at the back of the cell. The inlet tube is also thermally anchored to the 4\,K stage for precooling of helium to $\sim$5\,K before entering the cell. The precooled helium passes through a diffuser located 3\,mm from the gas inlet for good thermalization with the cell wall. The typical flow rate of helium is 15\,standard cubic centimeter per minute (sccm), and the cold helium collides with ablated CaOH molecules. The CaOH molecules are rapidly thermalized with helium and eventually exit the cell through the exit aperture, which is 5\,mm in diameter. Although we conducted experiments with different flow rates, difference were not observed in the results at the current measurement precision.

An ablation laser beam located 2.5\,cm from the exit aperture is sent through a hole at the center of the cell. A pulsed 532\,nm light from an Nd:YAG laser with $\sim$10\,ns width and $\sim$20\,mJ energy is used as the ablation laser. The ablation pulse is incident on a solid Ca(OH)$_2$ powder target inside the cell and produces CaOH molecules. With the current precision, the position where in the target the ablation laser points also does not change the results below.

Another hole with a 5-mm-diameter windows located 1\,cm from the exit aperture provides optical access to the absorption probe lasers. A ring-cavity dye laser (Coherent 899 dye laser, output power $\sim$500\,mW, bandwidth $\sim$1\,MHz) is used to excite the A--X transition of CaOH at approximately 625\,nm. The wavelength is monitored by a wavemeter (High finesse, WS6-200) with $\sim$200\,MHz absolute accuracy, which is not important in this measurement. Two counter-propagating absorption lasers with diameters of $\sim$1\,mm are injected perpendicular to the direction of the molecular flow. As shown in Fig.~\ref{fig:setup}, the laser along the ablation laser is represented by laser1, and the opposite-direction laser is denoted by laser2. Laser1 and laser2 have power of 20\,µW and 10\,µW, respectively.

We measured $Q_1(J=1/2)$ and $Q_1(3/2)$ transitions. The strong $Q_1(3/2)$ transition showed a good signal-to-noise ratio; however, it overlapped with the $R_{12}(1/2)$ transition owing to the spin--rotation interaction. In contrast, the $Q_1(1/2)$ transition appeared as an isolated peak. Therefore, the following sections mainly focus on the results of $Q_1(3/2)$, with those of $Q_1(1/2)$ being used for crosschecking. 

Based on Fig.~\ref{fig:setup}, two photodetectors are used to measure the probe laser signals, and another photodetector is placed to monitor the laser power before the laser enters the cell. The photodetector outputs were recorded by a four-channel oscilloscope (Tektronix, MSO64). Time trace of the laser transmittance over 20\,ms was recorded at a 250\,kHz sampling rate. Spectra were obtained by sweeping the probe laser frequency by $\sim$1.0\,GHz in 15\,s. The trigger of the oscilloscope was synchronized to the ablation laser at 10\,Hz.

\section{Results}

\begin{figure}[h]
\centering
\includegraphics[height=6.0cm]{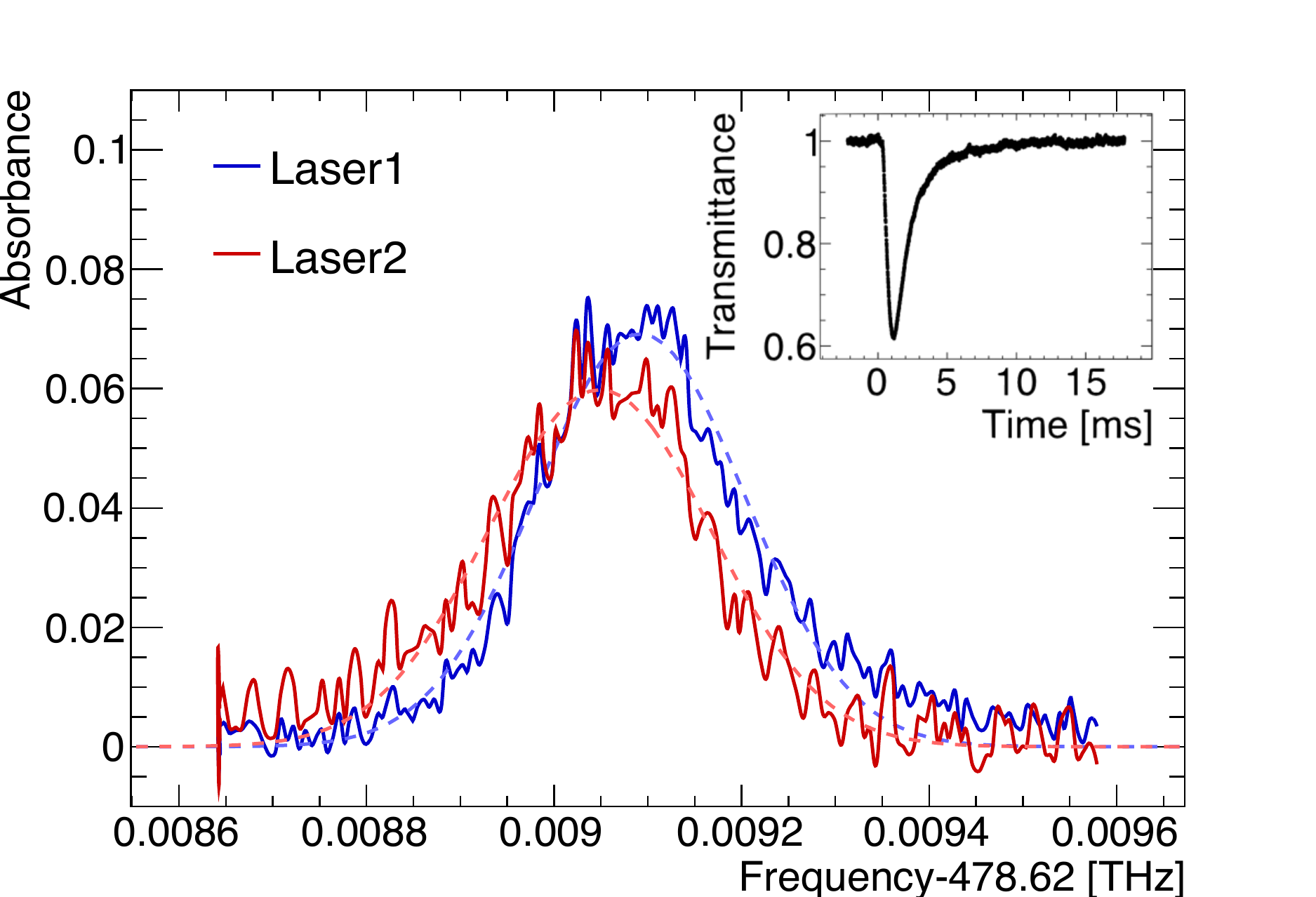}
\caption{\label{fig:spectrum}Example of transmittance trace (inset) and observed spectra of Doppler-limited absorption of CaOH $Q_1(3/2)$ transition probed by counter-propagating lasers at 0.3\,ms after ablation. Dashed lines correspond to Gaussian fitting results.}
\end{figure}

The inset in Fig.~\ref{fig:spectrum} shows an example transmittance trace over 20\,ms. The time origin corresponds to the ablation pulse timing. An absorbance spectrum was obtained by integrating the transmittance trace over 100\,ns at a certain delay time from ablation followed by normalization using the power monitor intensity. Figure~\ref{fig:spectrum} shows the spectra of the $J=3/2$ transition measured using the two counter-propagating lasers at a 0.3\,ms delay. The spectra were Gaussian fitted, and width and the peak frequency of each spectrum were obtained. The translational temperature of CaOH was estimated from the widths of the spectra. In Fig.~\ref{fig:spectrum}, both spectra have consistent widths that correspond to $\sim$38\,K. However, the peak frequencies show clear deviations, where the laser1 peak frequency is approximately 40\,MHz higher than the laser2 peak frequency. Here, 40\,MHz is a typical value, and the laser1 and laser2 peaks show anti-correlation over multiple datasets. This result indicates a Doppler shift, and that CaOH molecules have velocity components in the same direction as the laser1 propagation. 

\begin{figure}[h]
\centering
\includegraphics[height=6.0cm]{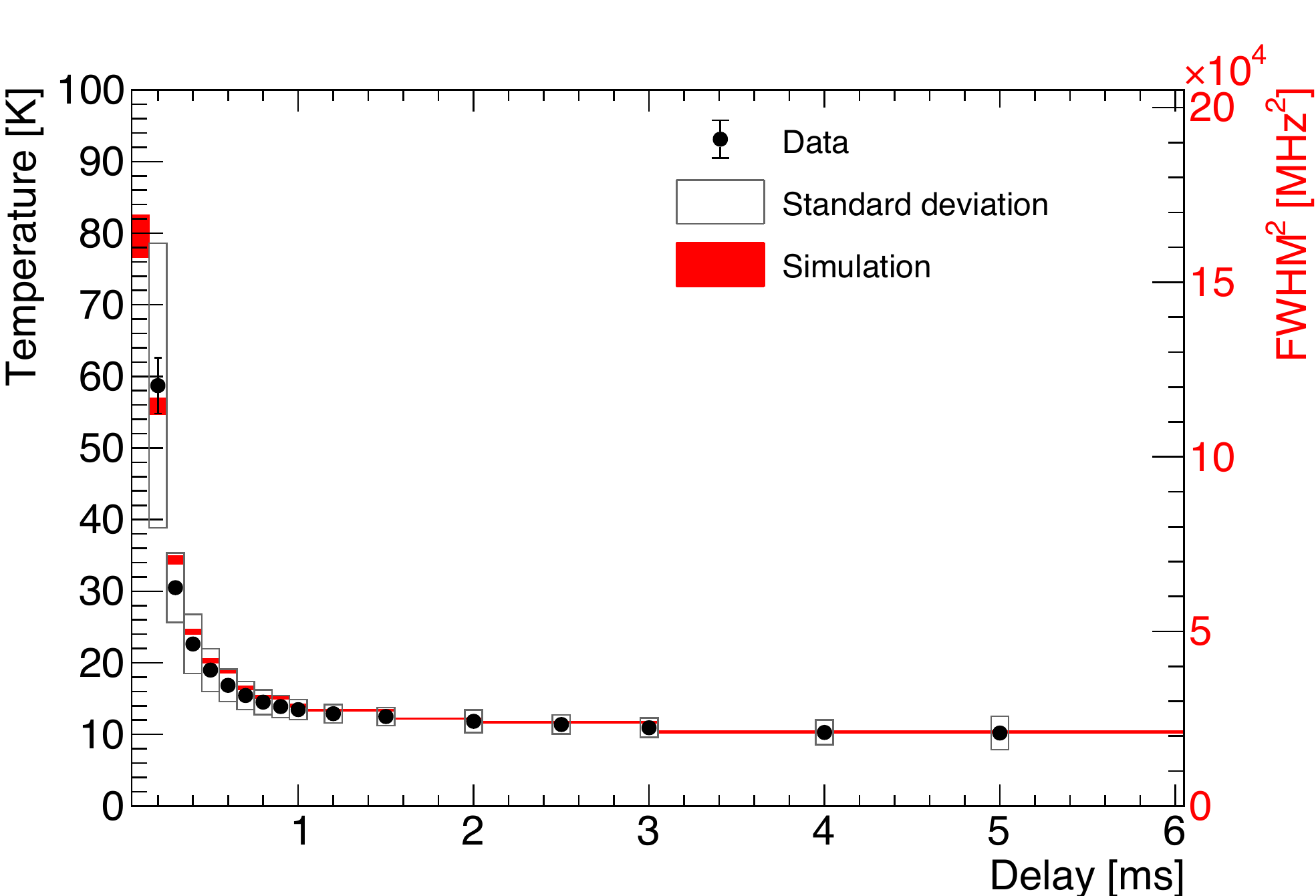}
\caption{\label{fig:sim1}Time evolution of CaOH translational temperature estimated from the Doppler width of absorption spectra of $Q_1(3/2)$ transition. The error bar on each data point corresponds to the standard error of multiple scan results, and the hatched region corresponds to that of standard deviation. The red band shows simulated result assuming that the helium temperature evolution follows the fit parameters obtained from the data. The right axis corresponds to the square of the full width at half maximum of the spectra.}
\end{figure}

The dependences of the Doppler width and shift on the time from the ablation are obtained by integrating the spectra at different delay times. Figure~\ref{fig:sim1} shows the time evolution of the temperature measured from the Doppler width of the $J=3/2$ transition. The spectrum was scanned approximately 20 times for each transition. In the figure, the error bar on each data point corresponds to the standard error of the multiple scan results, and the hatched region corresponds to that of the standard deviation. The Gaussian fit error of the spectrum is approximately a few MHz and not included in the plot. A time variance is clearly observable, where the temperature rapidly cools within $\sim$1\,ms and gradually dissipates over a few milliseconds. According to Ref.~\cite{Skoff2011}, the temperature evolution in a buffer-gas cell can be fitted with a double exponential function
\begin{equation}
\label{eq:1}
    T=T_c+T_1e^{-t/\tau_1}+T_2e^{-t/\tau_2}.
\end{equation}
The fitted results for $J=3/2$ are $T_c=10\pm1$ K, $T_1=158\pm44$ K, $T_2=7\pm1$ K, $\tau_1=0.13\pm0.02$\,ms, and $\tau_2=1.28\pm0.46$\,ms. The $J=1/2$ transition shows consistent results. Although the $R_{12}(1/2)$ transition broadens the $Q_1(3/2)$ width, this effect was not observable in this measurement. These results are also consistent with those reported in Ref.~\cite{Skoff2011} with the YbF molecule. The model in Ref.~\cite{Skoff2011} suggests that helium is heated by the ablation pulse; therefore, the temperature evolution in Fig.~\ref{fig:sim1} refers to the temperature of buffer-gas helium, which is dissipated with two different time constants, $\tau_1$ and $\tau_2$. 

To understand this time evolution process, we examined the CaOH thermalization with helium by performing a simple Monte Carlo simulation. The buffer-gas cell was constructed with the same dimensions in the simulation, and the measurement was reproduced in the following process. 

1) {\it CaOH initial state}: At the beginning of the simulation, CaOH molecules are spread out isotropically. Their initial velocity follows a Maxwell--Boltzmann distribution at 1000\,K. 

2) {\it Helium condition}: CaOH interacts with helium after passing through the mean free path, $l$. Here, we assume $l$ is a constant expressed as $l=1/(\sqrt{2}n_{He}\sigma)$, where $n_{He}$ is the helium density in the cell and $\sigma$ is the collisional cross-section between CaOH and helium. Helium density is estimated from the cell size and the typical flow (15\,sccm) as $n_{He}=5\times10^{15}/$cm$^3$. The collisional cross-section is assumed as $\sigma=3\times10^{-14}$\,cm$^2$. Here, no hydrodynamic calculation is implemented, and the helium gas simply has a 10\,m/s flow towards the cell aperture. The helium temperature is changed as follows the fitting result by Eq.~\ref{eq:1}. 

3) {\it Collision}: The hot CaOH molecules lose their energy by elastic hard-sphere collisions with cold helium buffer gas. The calculation is performed similarly to the description in Ref.~\cite{Takahashi2021}. The position and velocity of each CaOH molecule are tracked until they hit the cell walls. In this simulation, typically approximately 5\%  CaOH molecules reach the laser interaction region.

4) {\it Extraction of observables}: The temperature and the Doppler shift are calculated from the three-dimensional velocity of the CaOH molecules in the probe laser interaction region.

If the CaOH molecules are thermalized with helium well and the temperature evolution is in that of helium, the simulation result shows that the CaOH temperature dissipates with the same time constants as those for helium. The simulation result of the temperature evolution is shown in a red band in Fig.~\ref{fig:sim1}. Using the assumed helium density, CaOH molecules are rapidly thermalized with helium in under 0.1\,ms, and they cannot be accessed by experiments due to the low signal-to-noise ratio. In this simulation, we did not find any inconsistency with the model in Ref.~\cite{Skoff2011}. Conversely, the simulation with a constant helium temperature failed to reproduce the experimental results.

To examine the robustness of this simulation, we changed several initial parameters. First, the CaOH initial temperature was varied from 200\,K to 2000\,K. This change did not affect the simulation result, because the molecules were cooled sufficiently fast. Second, the CaOH initial emission angle was limited to forward (cos$\theta>$0.85), where $\theta$ is the angle between the emission direction and the axis perpendicular to the cell wall. The effect of this variation was negligible because the thermalization occurred spatially close to the ablation position, and the velocity distribution of the CaOH molecules became uniform direction after the thermalization. Finally, the effect of the mean free path, $l$, was investigated. The value $l$ was calculated from $l=1/(\sqrt{2}n_{He}\sigma)$, and we changed the helium density, $n_{He}$, from $5\times10^{14}/$cm$^3$ to $1\times10^{16}/$cm$^3$ in the simulation to vary the mean free path. This variation also corresponded to changing $\sigma$ from $3\times10^{-15}$cm$^2$ to $6\times10^{-14}$cm$^2$. In this range, the simulation results did not change. However, on applying a much lower helium density, the thermalization time constant increased depending on the other initial conditions. Contrastingly, the CaOH molecules with longer mean free paths collided with the cell walls more frequently; therefore, the survival rate of the CaOH molecules was significantly reduced in the simulation.

\begin{figure}[h]
\centering
\includegraphics[height=6.0cm]{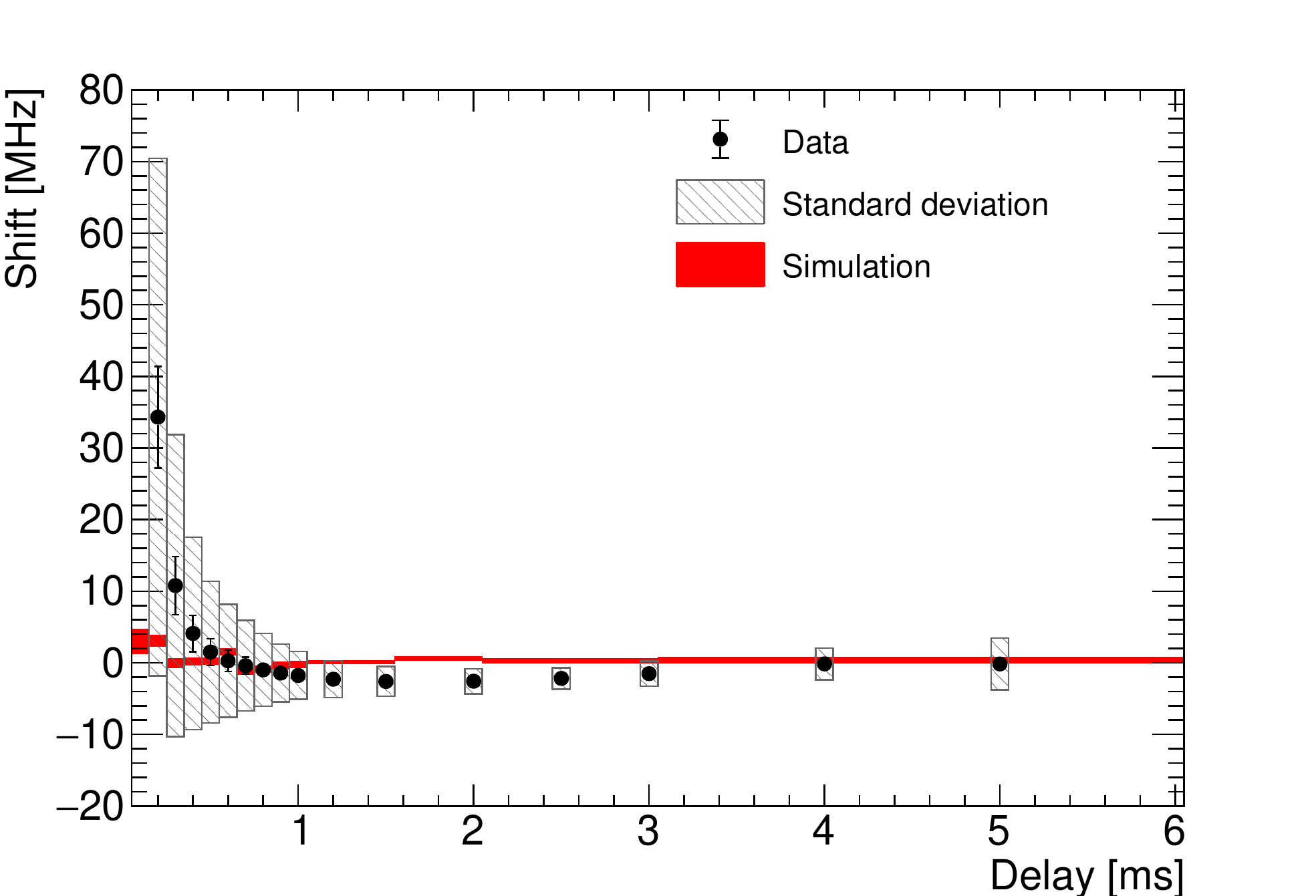}
\caption{\label{fig:sim2}Time evolution of observed Doppler shift. Error bar on data points, hatched region, and the red band correspond to the standard error of the multiple scan results, that of standard deviation, and simulation result, respectively.}
\end{figure}

\begin{figure}[h]
\centering
\includegraphics[height=6.0cm]{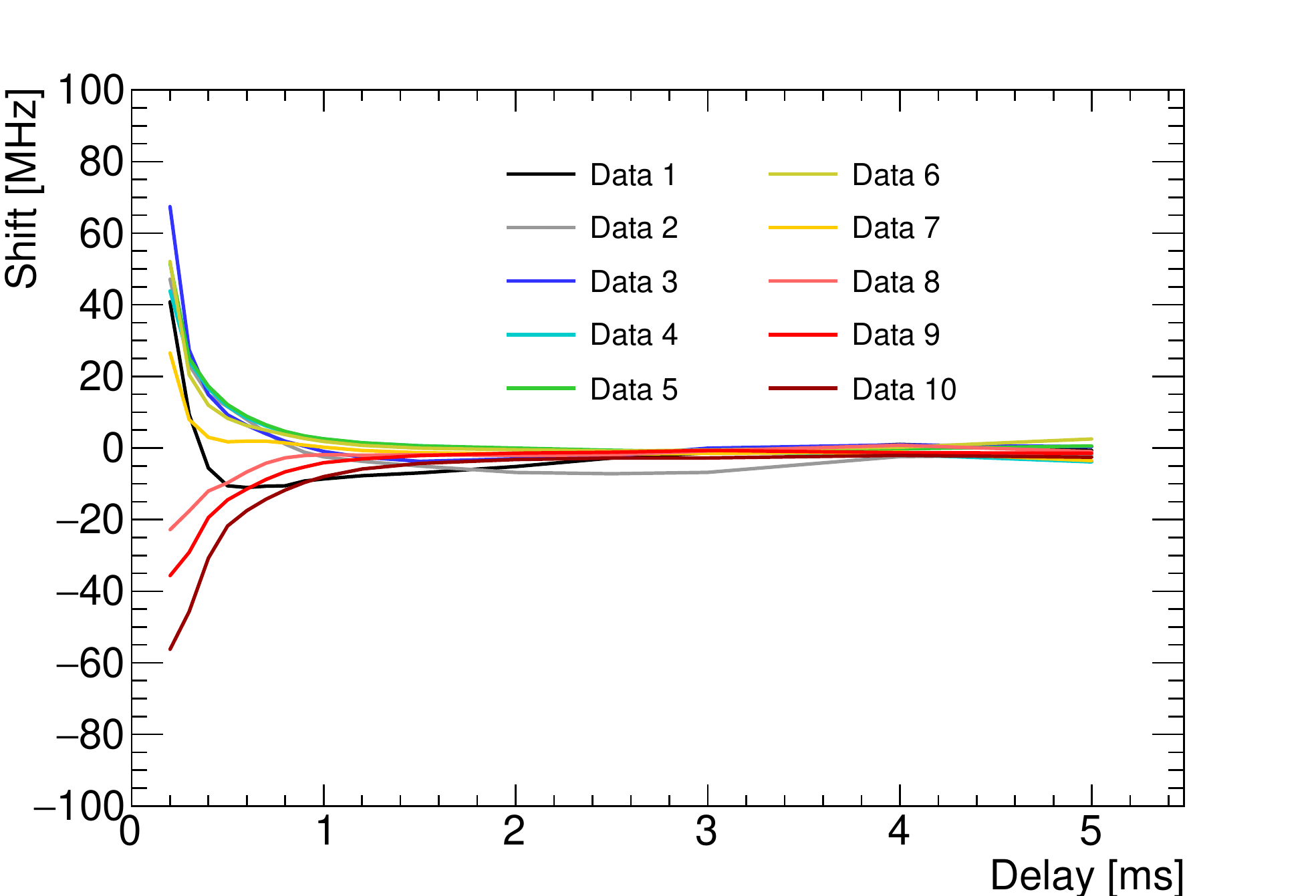}
\caption{\label{fig:shifteach}Time evolution of observed Doppler shift on multiple dataset. Data index is in order of data acquisition.}
\end{figure}

Figure~\ref{fig:sim2} shows the data and simulation results of the time evolution of the Doppler shift. Shifts in the data are obtained from the half of difference between the peak frequencies of laser1 and laser2. The shift depends on the delay time, particularly below 1\,ms. Here the error bar on the data point, hatched region, and red band correspond to the standard error of the multiple scan results, standard deviation, and simulation result, respectively. The standard deviation is much larger than the simple statistical fluctuation because the Doppler shift varies with the dataset. Figure~\ref{fig:shifteach} shows examples of time evolution of the Doppler shift before averaging over the entire dataset. Each line corresponds to a single dataset, and the data index is in order of data acquisition. Although most datasets show shifts in the same direction, several scans present even opposite shift trends. In addition, some neighboring data sets show similar shift trends. 

These results suggest that the velocity of CaOH changes its direction on a time scale of a few minutes. This change can be due to the complex flow of helium in the cell, which gradually changes its velocity and direction. Because the present simulation does not include such a complex flow, almost no Doppler shift occurs in the simulation results, as shown in Fig.~\ref{fig:sim2}. The large deviations shown experimentally can be explained by this gradually changing flow model. Helium can flow in the opposite direction than normal, in which case the sign of the Doppler shift changes. The origin of this flow may be helium convection, flow due to the diffuser, or leakage from the cell, although no specific answer is obtained.

%These results suggests that there is a helium flow such as convection, and inversion of the direction of flow propagation change the sign of Doppler shift. The large deviation in Fig.~\ref{fig:sim2} can be explained by variety of flow the helium velocity between data set. 

%On the other hand, we do not see any shift in the simulation in Fig.~\ref{fig:sim2} because the CaOH molecules are thermalized faster than the time scale shown in Fig.~\ref{fig:sim2}. This result implies that there is a flow caused by other effects than the thermalization, such as helium convection, helium flow caused by the diffuser, and helium leakage from the cell, which are not included in our simulation. 

\section{Conclusion}

In this study, we probed the Doppler effects in a buffer-gas cell by high-resolution absorption spectroscopy of CaOH molecules using two counter-propagating lasers. The translational temperature measured from the Doppler width showed that the target molecules were cooled within 1\,ms and gradually dissipated in a few milliseconds. This behavior was reproduced well in a Monte Carlo simulation using a previously proposed model, which included the effect of helium temperature. The Doppler shift also showed time evolution, which could not be explained by the model. We concluded that the shift can be caused by the complex flow in the cell, whose velocity and direction gradually vary; however, the origin of the flow was unclear. This assumption may explain the large variation in the Doppler shift experimentally observed. The Doppler shift was on the order of 10\,MHz after 1\,ms from the ablation, when a strong signal was expected. This 10\,MHz-order shifts cannot be ignored in high-resolution spectroscopy and must be treated carefully. The best solution to address this problem is Doppler-free spectroscopy. However, this method may result in a small signal-to-noise ratio. Optimizing the cell design may solve this problem.

\begin{acknowledgments}

We would like to thank the members
of Core for Quantum Universe (RIIS, Okayama University). Y. T. would like to thank the Masason Foundation
for their financial support. This work was supported by
JSPS KAKENHI Grant Nos. 18H01229 and 22H01249,
and Masason Foundation. We would like to thank Editage (www.editage.com) for English language editing.

\end{acknowledgments}

\bibliography{apssamp}% Produces the bibliography via BibTeX.

\appendix

\end{document}